\documentclass[11pt,a4paper]{article}
\usepackage[a4paper,margin=27mm,headheight=15pt]{geometry}
\usepackage[T1]{fontenc}
\usepackage[utf8]{inputenc}
\usepackage{amsmath,amssymb,amsthm,mathtools}
\usepackage{xcolor}
\IfFileExists{newtxtext.sty}{\usepackage{newtxtext,newtxmath}}{\usepackage{lmodern}}
\usepackage{microtype}
\usepackage{booktabs,array,longtable,enumitem}
\usepackage{fancyhdr}
\usepackage[hidelinks,pdfencoding=auto,psdextra]{hyperref}
\usepackage[nameinlink,noabbrev]{cleveref}
\definecolor{CASgreen}{RGB}{34,105,68}
\newcommand{\CASmark}[1]{\ensuremath{\;{\color{CASgreen}\clubsuit_{\scriptscriptstyle\mathrm{#1}}}}}
\newcommand{\E}{\mathbb E}
\newcommand{\R}{\mathbb R}
\newcommand{\Pp}{\mathbb P}

\newcommand{\aZero}{\alpha_0}
\newcommand{\rLow}{\rho_{\mathrm L}}
\newcommand{\rHigh}{\rho_{\mathrm H}}
\newcommand{\G}{\mathcal G}
\newtheorem{theorem}{Theorem}
\newtheorem{lemma}{Lemma}
\newtheorem{proposition}{Proposition}

\theoremstyle{definition}

\newtheorem{conjecture}{Conjecture}
\theoremstyle{remark}

\allowdisplaybreaks[2]
\hypersetup{pdftitle={Noise Stability on the Hypercube: Order 7/8},pdfauthor={Lei Yu},pdfsubject={Balanced Boolean functions, asymmetric power stability, and Hellinger-kernel transfer}}

\begin{document}
\title{Noise Stability on the Hypercube: Order $7/8$}
\author{Lei Yu\thanks{L. Yu is with the School of Statistics and Data Science, LPMC, KLMDASR,
and LEBPS, Nankai University, Tianjin 300071, China (e-mail: leiyu@nankai.edu.cn).}}
\date{}
\maketitle

\begin{abstract}
For a balanced Boolean function $f$ on the discrete hypercube, we study the \(\alpha\)-stability $\E(T_\rho f)^\alpha$, where $T_\rho$ is the Boolean noise operator and $0<\alpha<1$. Using the asymmetric Hellinger kernel estimates recently established by OpenAI, we prove that dictators minimize the \(7/8\)-stability in every dimension and at every correlation $0\le\rho\le1$. At interior correlations they are the only minimizers. As a consequence, dictators minimize the \(\alpha\)-stability for every  $7/8\le\alpha<1$, maximize the \(\alpha\)-stability for every  $1<\alpha<2$, and  maximize the one-sided Shannon entropy stability $\E[T_\rho f\log T_\rho f]$. This resolves a conjecture of Li and M\'edard and strengthens the Courtade--Kumar conjecture in the balanced setting.
\end{abstract}


\section{Introduction}\label{sec:intro}
Let $X=(X_1,\ldots,X_n)$ be uniformly distributed on $\{-1,1\}^n$. For $0\le\rho\le1$, obtain $Y$ from $X$ by independently flipping each coordinate with probability $p=(1-\rho)/2$. The Bonami--Beckner noise operator is
\begin{equation}\label{eq:noise}
 T_\rho f(x)=\E[f(Y)\mid X=x].
\end{equation}
For a Boolean indicator $f:\{-1,1\}^n\to\{0,1\}$,   $T_\rho f(x)$ is the conditional probability of the event $f(Y)=1$. A Boolean indicator $f$ (or the sign function $g=2f-1$)  is \emph{balanced} if $\E f=1/2$. 
A signed coordinate indicator $d(x)=(1\pm x_i)/2$ (or $g(x)=\pm x_i$) is called a \emph{dictator}.

The quantity $\E[(T_\rho f)^\alpha]$ is known as the $\alpha$-stability of $f$. The $\alpha$-stability problem, in a general sense, concerns which Boolean
functions (or measurable sets) attain the extremum of the $\alpha$-stability. In other words, which  Boolean
functions are the ``most stable'' under the
action of the noise operator in terms of the $\alpha$-norm. 

We write
\begin{equation}\label{eq:dictator-benchmark}
 \bar{p}=1-p=\frac{1+\rho}{2},\qquad
 M_\alpha(\rho)=\frac{p^\alpha+\bar{p}^\alpha}{2}.
\end{equation}
For a signed coordinate indicator $d(x)=(1\pm x_i)/2$, usually called a dictator, $T_\rho d$ takes the values $p,\bar{p}$ with equal probabilities. Consequently $\E[(T_\rho d)^\alpha]=M_\alpha(\rho)$.

The extremal problem considered here is the following two conjectures. 
\begin{conjecture}[$1/2$-stability conjecture \cite{Yu}]\label{conj:main}
For every $1/2\le\alpha<1$, every correlation, and every balanced Boolean functions $f$,
\begin{equation}\label{eq:target}
 \E[(T_\rho f)^\alpha] \ge  M_\alpha(\rho).
\end{equation}
\end{conjecture}
This is a strengthening of the Li--M\'edard conjecture \cite{LiMedard}. 
\begin{conjecture}[$\alpha$-stability conjecture for $\alpha \in (1,2)$ \cite{LiMedard}]\label{conj:main2}
Inequality \eqref{eq:target} with reverse direction holds for every $1<\alpha<2$, every correlation, and every balanced Boolean functions.
\end{conjecture}

Define the $\alpha$-Tsallis kernel
\begin{equation}\label{eq:tsallis}
H_\alpha(t)=\frac{t^\alpha-t}{\alpha-1},
\end{equation}
and the (one-sided) $\alpha$-Tsallis stability as 
$\E[H_\alpha(T_\rho f)]$. 
The affine term has fixed expectation and the denominator is positive for $\alpha>1$ and negative for $\alpha<1$. 
Thus, the two conjectures above indicate that dictators maximize $\alpha$-Tsallis stability for $\alpha \in [1/2,1)\cup (1,2)$. 
Moreover, $H_\alpha(t)\to t\log t$ as $\alpha \to 1$, with $0\log0=0$. 
All logarithms in this paper take the natural base $e$. 
Thus the two conjectures above for exponents approaching one yields a one-sided Shannon entropy inequality.

\subsection{Main results and proof strategy}
Our main result is the resolution of a weaker version of \cref{conj:main} and the original \cref{conj:main2}. 
\begin{theorem}[$7/8$-stability]\label{thm:main}
Assume \cref{ass:input}. For every $n\ge1$, every balanced Boolean indicator $f:\{-1,1\}^n\to\{0,1\}$, every $7/8\le\alpha<1$, and every $0\le\rho\le1$,
\begin{equation}\label{eq:main}
 \E(T_\rho f)^\alpha\ge M_\alpha(\rho).
\end{equation}
If $0<\rho<1$, equality holds if and only if $f(x)=(1\pm x_i)/2$ for some coordinate $i$. 
In particular, \cref{conj:main2} holds and 
\begin{equation}\label{eq:Shannon}
 \E[T_\rho f\log T_\rho f]\le\frac{p\log p+\bar{p}\log \bar{p}}{2}.
\end{equation} 
\end{theorem}

Throughout this paper, we set
\begin{equation}\label{eq:thresholds}
 \aZero=\frac78,\qquad
 \rLow=\frac{543}{1000},\qquad
 \rHigh=\frac{\sqrt{184}}{25}.\CASmark{01}
\end{equation}
The proof consists of two principal components. 
First, by the same strategy developed in \cite{YuPhi,Yu}, reverse hypercontractivity and an explicit first-level Fourier bound are applied to prove the base exponent $7/8$ for $\rho\le\rLow$. Second, another double-contact minorant transfers the one-sided Hellinger-information inequality from \cite{OpenAI} to the base exponent for $\rho\ge\rHigh$. The two ranges overlap, since $\rLow>\rHigh$.

The base exponent $7/8$ has not been optimized and may be
further lowered within the present proof framework.
However, reaching the critical exponent $1/2$ appears
to require a new auxiliary kernel for the induction argument.

\paragraph{Mathematica verification marks.}
The proofs in this paper are analytic, although some supporting
symbolic calculations are lengthy. To facilitate independent
verification, we provide accompanying Mathematica code for these
calculations. A green clover symbol \CASmark{01}, with a unique
serial number, marks each formula, coefficient table, or fixed-value
comparison covered by the code. Each mark corresponds to a
verification routine with the matching label
\texttt{CAS-01}, etc. A complete cross-reference is provided in
Appendix~\ref{app:CAS-index}.
These marks indicate checks of exact symbolic identities,
coefficient inequalities, or rigorously certified fixed-value
bounds. They do not constitute machine verification of the
analytic proofs or of the external hypothesis in \cref{ass:input}.



\subsection{Motivations and related works}
\label{subsec:motivations}

Noise stability of Boolean functions is closely connected to
isoperimetric inequalities, hypercontractivity, and Fourier
analysis on the discrete hypercube. For a Boolean function
$f=\mathbf{1}_A$, the noise operator $T_\rho f$ may be viewed
as a smoothed indicator of the set $A$. Extremal inequalities
for functionals of $T_\rho f$ can therefore be interpreted
as nonlinear or ``soft-boundary'' versions of isoperimetric
inequalities. In the classical quadratic setting, the
edge-isoperimetric problem on the hypercube can be recovered
by considering the limit $\rho\to1$
\cite{ledoux1994semigroup,CK,LiMedard,Yu}.
For $0<\alpha<1$, the functional
$\mathbb E(T_\rho f)^\alpha$ leads to a different soft-boundary
quantity. Indeed, writing $p=(1-\rho)/2$ and letting
$d_A(x)$ denote the number of neighbors of $x$ belonging
to $A$, one has \cite{Yu}, for fixed dimension, 
\[
\mathbb E(T_{1-2p}\mathbf{1}_A)^\alpha
=
\mathbb P(A)
+
p^\alpha\mathbb E\big[
\mathbf{1}_{A^c}(X)d_A(X)^\alpha
\big]
+o(p^\alpha),
\qquad p\downarrow0.
\]
Thus the subunit-order problem retains a geometric
interpretation, but its boundary contribution differs from
the classical edge boundary. Noise stability inequalities
are also closely related to hypercontractivity and
semigroup inequalities
\cite{nair2014equivalent,ledoux2014remarks}.
Furthermore, the classical quadratic noise stability
admits the Fourier representation
\[
\mathbb E[fT_\rho f]
=
\sum_{S\subseteq[n]}\rho^{|S|}\widehat f(S)^2,
\]
which is the generating function of the Fourier weights
of $f$ \cite{ODonnell}. These connections motivate the
study of general nonlinear noise-stability functionals.

The study of related extremal problems dates back to the
seminal works of G\'acs and K\"orner
\cite{gacs1973common} and Witsenhausen
\cite{witsenhausen1975sequences}, particularly in connection
with non-interactive correlation distillation (NICD).
For balanced Boolean functions, Witsenhausen's
maximal-correlation argument establishes the optimality of
dictators for the quadratic stability problem. 
Mossel and O'Donnell \cite{mossel2005coin} subsequently
studied higher-order symmetric (two-sided)  stability $\mathbb E[
(T_\rho f)^\alpha+(1-T_\rho f)^\alpha]$ and proved
dictator optimality at order $3$. This  was extended by
the present author \cite{YuPhi}, who established dictator
optimality for asymmetric orders in $[2,3]$ and symmetric
orders in $[2,5]$, respectively. On the other hand,
dictator optimality fails in general at order $10$, as
observed by Mossel and O'Donnell. These results illustrate
the importance of the exponent in determining the
extremizers of noise-stability functionals. 

Another central direction is the Courtade--Kumar problem
\cite{CK}, which concerns the maximization of mutual
information between a Boolean function of noisy observations
and the original input. Samorodnitsky
\cite{samorodnitsky2016entropy} established a
dimension-independent low-correlation range for this
conjecture. The present author \cite{YuPhi} subsequently
obtained an explicit correlation threshold of approximately
$0.46$ in the balanced setting, together with a finite-dimensional optimization
bound, and then  \cite{Yu} improved the correlation threshold to 
$0.914$ by a local optimality argument.
More recently, the Courtade--Kumar conjecture has
been resolved in full generality by several groups using different methods
\cite{Chen,KT,MB,KramerSaglam,OpenAI2}. 
Li and M\'edard  \cite{LiMedard} investigated the asymmetric (one-sided) stability for order in $(1,2)$, related this quantity  to the one-sided Shannon information, 
and posed Conjecture \ref{conj:main} as a strengthening of the Courtade--Kumar conjecture in the balanced setting. 
As a strengthening of the Courtade--Kumar conjecture, 
Anantharam et al.\ \cite{ABCJN} posed the Hellinger-information conjecture in which the Shannon mutual information is replaced by the Hellinger-information. Now this conjecture has  
been resolved by OpenAI as well \cite{OpenAI}.
Other related results include
the two-function mutual-information inequality of
Pichler, Piantanida, and Matz
\cite{pichler2018dictator}, and the
Gaussian-profile approach of Chen and Nair \cite{CN}.
In Gaussian space, Borell's isoperimetric theorem
\cite{borell1985geometric} identifies half-spaces as
extremizers of classical noise stability; related
functional inequalities were also investigated in
\cite{kindler2015remarks,eldan2015two}.
Recent sharp isoperimetric inequalities on the discrete
cube \cite{DIRX} provide further connections between
noise operators and geometric inequalities.

The present paper concerns the asymmetric (one-sided) $\alpha$-stability
problem in the previously less understood range
$0<\alpha<1$. In this range, minimizing
$\mathbb E(T_\rho f)^\alpha$ among Boolean functions of
fixed mean is equivalent to maximizing the corresponding
$\alpha$-Tsallis stability. Unlike the symmetric functional
\[
\mathbb E\big[
(T_\rho f)^\alpha+(1-T_\rho f)^\alpha
\big],
\]
the asymmetric functional retains information about the
distribution of $T_\rho f$ that is lost under symmetrization.
Consequently, symmetric stability inequalities do not
directly imply their asymmetric counterparts.

\paragraph{Organization.}
Section~\ref{sec:prelim} introduces the notation and preliminary
results. Sections~\ref{sec:low} and~\ref{sec:high} establish the
main inequality in the low- and high-correlation regimes,
respectively. Section~\ref{sec:completion} completes the proof
and derives its consequences. The appendices provide detailed
proofs of the auxiliary results using elementary calculus,
exact polynomial identities, and Taylor expansions with
Cauchy remainder bounds.

\section{Notation and preliminary facts}\label{sec:prelim}
All expectations on a cube use normalized uniform measure. Given an indicator $f$, write $g=2f-1$. For $S\subseteq[n]$, let $\chi_S(x)=\prod_{i\in S}x_i$. Then
\begin{equation}\label{eq:fourier-expansion}
 \widehat g(S)=\E[g\chi_S],\qquad
 g=\sum_{S\subseteq[n]}\widehat g(S)\chi_S,\qquad
 T_\rho g=\sum_{S\subseteq[n]}\rho^{|S|}\widehat g(S)\chi_S.
\end{equation}
Set $W_k(g)=\sum_{|S|=k}\widehat g(S)^2$. For a balanced Boolean sign function,
\begin{equation}\label{eq:fourier-basic}
 W_0=0,\qquad \sum_{k\ge1}W_k=1,\qquad
 \E[gT_\rho g]=\sum_{k\ge1}\rho^kW_k,\qquad
 \E(T_\rho g)^2=\sum_{k\ge1}\rho^{2k}W_k\le\rho^2.
\end{equation}
The distance to the closest signed dictator is
\begin{equation}\label{eq:distance}
 \delta(g)=\min_{i,\,\sigma\in\{-1,1\}}
 \Pp(g(X)\ne\sigma X_i)
 =\frac{1-\max_i|\widehat g(\{i\})|}{2}.
\end{equation}
For $0<\rho<1$ and a nonconstant Boolean indicator, the noise kernel is strictly positive, and hence $0<T_\rho f<1$ at every point.

\paragraph{Upward exponent transfers.}  By upward exponent transfers in the following proposition, it remains to establish the base exponent $\aZero=7/8$. 

\begin{proposition}[Upward exponent transfers \cite{BarnesOzgur,Yu}]\label{cor:up}
For $0<a<b<2$, if dictators maximize $a$-Tsallis stability, then dictators maximize $b$-Tsallis stability. 
\end{proposition}

\paragraph{Coordinate sorting.}
For every continuous concave $K:[0,1]\to\R$ and Boolean indicator $f$, coordinate sorting gives an increasing Boolean $f^\uparrow$ with the same mean and
\begin{equation}\label{eq:sorting}
 \E K(T_\rho f)\ge\E K(T_\rho f^\uparrow).
\end{equation}
This familiar rearrangement follows by replacing the two sections in each coordinate by their pointwise minimum and maximum; see \cite{CK,LiMedard,Yu}. In particular, we will apply it to the concave kernels constructed in \cref{sec:high}.

\paragraph{Reverse hypercontractivity.}
We use the standard uniform-cube inequality \cite[Eq.~(1.2)]{MOS}
\begin{equation}\label{eq:rhc}
 \|T_\rho v\|_b\ge\|v\|_a,\qquad b<a<1,\quad
 \rho^2\le\frac{1-a}{1-b},\quad v>0.
\end{equation}
Negative exponents are allowed; nonnegative inputs follow by a positive approximation. This is the normalization underlying the section estimates of \cite{Yu}.

\paragraph{Asymmetric Hellinger kernel bounds.}
Define the Hellinger profile on $[0,1]$ by
\begin{equation}\label{eq:hellinger-profile}
 h(t)=2\sqrt{t(1-t)},\qquad s=\sqrt{1-\rho^2}=h(p).
\end{equation}
For a polynomial $\lambda$, put
\begin{equation}\label{eq:source-kernel}
 \G_\lambda(t)=h(t)\left(1+
       \frac{2t-1}{1+h(t)\lambda(h(t))}\right).
\end{equation}
The denominators below are positive because each listed quadratic is positive on $[0,1]$.

\begin{proposition}[Asymmetric  Hellinger kernel bounds \cite{OpenAI}]\label{ass:input}
For every increasing balanced Boolean indicator $f$,
\begin{equation}\label{eq:source-bound}
 \E\G_{\lambda_j}(T_\rho f)\ge s
\end{equation}
whenever $s$ lies in the corresponding band in \cref{tab:source}. At a common endpoint either of the indicated kernels may be used.
\end{proposition}
Proposition \ref{ass:input} is the balanced specialization of Proposition~3.5 and Table~1 of  \cite{OpenAI}, using the normalization $B(0)=1$ in its Lemma~3.4. 

\begin{table}[htbp]
\centering
\caption{The three asymmetric Hellinger kernels. All coefficients are exact.}\label{tab:source}
\begin{tabular}{@{}ccl@{}}\toprule
$j$ & $s$ interval & $\lambda_j(x)$ \\\midrule
$1$ & $(0,4/25]$ & $\displaystyle \frac{47}{50}-\frac{11}{25}x+\frac3{10}x^2$ \\[5pt]
$2$ & $[4/25,3/10]$ & $\displaystyle \frac{443}{500}-\frac{199}{500}x+\frac14x^2$ \\[5pt]
$3$ & $[3/10,21/25]$ & $\displaystyle \frac{421}{500}-\frac{67}{200}x+\frac9{50}x^2$ \\\bottomrule
\end{tabular}
\end{table}

\section{The low-correlation theorem}\label{sec:low}
Throughout this section set $\alpha=\aZero=7/8$, $g=2f-1$, and $U=T_\rho g$. Define the convex function
\begin{equation}\label{eq:normalized-Phi}
 \Phi(u)=\frac{1+\alpha u-(1+u)^\alpha}{\alpha(1-\alpha)},
 \qquad
 \Psi(r)=\frac{\Phi(r)+\Phi(-r)}2.
\end{equation}
Since $\E U=0$ and $T_\rho f=(1+U)/2$, the desired inequality is equivalent to
\begin{equation}\label{eq:Phi-target}
 \E\Phi(U)\le\Psi(\rho).
\end{equation}
The normalization gives
\begin{equation}\label{eq:Phi-curvature}
 \Phi''(u)=(1+u)^{-9/8},\qquad
 \Psi''(r)=\frac{(1+r)^{-9/8}+(1-r)^{-9/8}}2\ge1.\CASmark{02}
\end{equation}
As $\Psi(0)=\Psi'(0)=0$, it follows that $\Psi(r)\ge r^2/2$.

We divide the proof according to the largest first-level Fourier coefficient. A direct reverse-hypercontractive estimate treats a neighborhood of a dictator, which is due to the author  \cite{Yu}. 
Outside that neighborhood, based on  the author's framework  \cite{YuPhi,Yu}, a polynomial majorant (i.e., polynomial upper bound) for the absolute value gives the first-level bound needed for the remaining moment estimates.

\subsection{A dictator neighborhood of radius \texorpdfstring{$27/100$}{27/100}}\label{sec:local}
Let $m=\max_i|\widehat g(\{i\})|$ and $\delta=(1-m)/2$. Choose a maximizing coordinate and reverse its sign if necessary so that its coefficient is positive. The sections $f_+,f_-$ in this coordinate have means $1-\delta$ and $\delta$.

We now need the reverse hypercontractivity bound for  $q$-stability due to the author  \cite[Theorem~4.6]{Yu}.

\begin{proposition}[Reverse hypercontractivity bound on  $\alpha$-stability  \cite{Yu}]
\label{prop:qstability} For any balanced Boolean function $f$ and  for any $0<\alpha<1$,  it holds that 
\begin{equation}
\E(T_\rho f)^{\alpha} \ge  \frac{1}{2}[(\delta+ (1-2\delta) \bar{p}^{\beta})^{\alpha/\beta}+ (\delta+(1-2\delta) p^{\beta} )^{\alpha/\beta}] \label{eq:HCbound}
\end{equation}
with $\beta=1+(\alpha-1)\rho^{2}$. 
\end{proposition} 

Applying  this proposition  with output exponent $\alpha=7/8$ and input exponent $\beta=1-\rho^2/8$ gives
\begin{equation}\label{eq:local-L}
 \E(T_\rho f)^{7/8}\ge L_\rho(\delta)
 :=\frac12\left[
 (\delta+(1-2\delta)\bar{p}^{\beta})^{7/(8 {\beta})}+
 (\delta+(1-2\delta)p^{\beta})^{7/(8 {\beta})}\right].
\end{equation}
The right side of \eqref{eq:local-L} is concave in $\delta$ and satisfies $L_\rho(0)=M_{7/8}(\rho)$. When it exists in $(0,1/2)$, the nonzero endpoint of this local comparison is the unique positive solution of $L_\rho(\delta)=M_{7/8}(\rho)$ \cite{Yu}. The argument below proves directly that the comparison holds on $[0,27/100]$ throughout the required correlation range. 

We now provide a universal  threshold $\delta^*$ for all  $0<\rho\le\rLow$. 
Define 
\begin{equation}\label{eq:local-D}
 D(\rho):=L_\rho(27/100)-M_{7/8}(\rho).
\end{equation}
The following analytic comparison proves the required uniform bound.
Let $\ell=355/512>\log2$, $a=7/8$, $d=27/100$, $m=23/50$, and define
\begin{equation}\label{eq:local-analytic-branch}
 \mathcal J(z)=
 \left(2d e^{-\ell z^2/8}+m(1+z)^{1-z^2/8}\right)^{a/(1-z^2/8)}
 -(1+z)^a.
\end{equation}
Near zero use the analytic branches with value $1$ for the powered
terms, and set
\begin{equation}\label{eq:local-analytic-F}
 \mathcal F(v)=\frac{\mathcal J(\sqrt v)+\mathcal J(-\sqrt v)}2.
\end{equation}
Since replacing $\log2$ by $\ell$ decreases each
positive summand of the local bound,
\[
 D(\rho)\ge 2^{-7/8}\mathcal F(\rho^2).
\]
In Appendix~\ref{app:local-interval}, using a Taylor expansion and
Cauchy's coefficient estimate, we  prove the following lemma.  
\begin{lemma}\label{lem:local-derivatives} 
It holds that
\begin{equation}\label{eq:local-derivatives} 
 \mathcal F''(v)<0\quad(0\le v\le\rLow^2),\qquad
 \frac{2^{-7/8}\mathcal F(\rLow^2)}{\rLow^2}>\frac1{25000}.
\end{equation}
\end{lemma}

The second inequality is a comparison of a single explicit constant.
As $\mathcal F(0)=0$, concavity gives
\begin{equation}\label{eq:local-gap}
 D(\rho)\ge
  2^{-7/8}\mathcal F(\rho^2)
 \ge
 \frac{\rho^2}{\rLow^2}\,2^{-7/8}\mathcal F(\rLow^2)
 >\frac{\rho^2}{25000}\qquad(0<\rho\le\rLow).
\end{equation}
We have  $ L_\rho(0)=M_{7/8}(\rho)$ and $L_\rho(27/100)>M_{7/8}(\rho)$. Concavity $L_\rho(\delta)$ in $\delta$ then proves $ L_\rho(\delta) \ge M_{7/8}(\rho)$ whenever $\delta\le27/100$. Combining this with \eqref{eq:local-L} gives \eqref{eq:target} and the inequality in  \eqref{eq:target} holds strictly except for dictators at positive correlations.

\subsection{A quantitative first-level Fourier bound}\label{sec:fourier-bound}
The nonlocal argument requires more than Parseval's identity. The next lemma supplies an explicit estimate at the coefficient threshold complementary to the local estimate above.
The proof is   given in
Appendix~\ref{app:W1-proof}. 

\begin{lemma}[First-level weight]\label{lem:W1}
If a (not necessarily balanced) Boolean function $g:\{-1,1\}^n\to\{-1,1\}$ satisfies
\[
 \max_i|\widehat g(\{i\})|\le\frac{23}{50},
\]
then
\begin{equation}\label{eq:W1-bound}
 W_1(g)\le\frac34.
\end{equation}
\end{lemma}

\subsection{Moment bounds retaining the Boolean input}\label{sec:source-moments}
For the rest of the low-correlation proof suppose $m\le23/50$ and put $w=3/4$. The previous lemma gives $W_1\le w$. The variable $U$ is not treated as an arbitrary centered random variable: the correlation between $U=T_\rho g$ and its Boolean source $g$ yields useful extra constraints.

The Fourier expansion gives the two estimates we need:
\begin{align}
 \E U^2&\le\rho^2\{w+(1-w)\rho^2\},\label{eq:moments-three}\\
 a\E U^2-b\E[gU]&\le w(a\rho^2-b\rho),
 \qquad a\rho^2\le b\le a\rho.\label{eq:mixed}
\end{align}
For the first, isolate Fourier level one and use
$\rho^{2k}\le\rho^4$ for $k\ge2$. For the second, the exact expansion is
\[
 a\E U^2-b\E[gU]
 =(a\rho^2-b\rho)W_1+
   \sum_{k\ge2}\rho^k(a\rho^k-b)W_k.
\]
The first coefficient is nonnegative and every summand in the tail is
nonpositive. Thus $W_1\le w$ proves \eqref{eq:mixed} directly.

For a negative exponent $\gamma<0$ put
$\beta_\gamma=1+(\gamma-1)\rho^2>0$.
Reverse hypercontractivity applied to $1+g=2f$ gives
$\|1+U\|_\gamma\ge2^{1-1/\beta_\gamma}$.
Raising to the negative power $\gamma$ reverses the inequality, so
\begin{equation}\label{eq:negative-moment}
 \E(1+U)^\gamma\le2^{\gamma-\gamma/\beta_\gamma}.
\end{equation}
Positive approximation handles zero values of the input.
For $\gamma=-1/2$ write
\begin{equation}\label{eq:B-negative}
 B(\rho)=2^{3\rho^2/(4-6\rho^2)}-1.
\end{equation}

\subsection{The interval \texorpdfstring{$0<\rho\le1/10$}{0 < rho <= 1/10}}
The following global scalar bound avoids the need for a certificate near the zero-correlation equality point:
\begin{equation}\label{eq:small-scalar}
 \Phi(u)\le\frac{23}{64}u^2+
 \frac3{64}\big((1+u)^{-2}-1+2u\big),\qquad u>-1.\CASmark{03}
\end{equation}
Indeed, the residual has value and first derivative zero at $u=0$, while its second derivative equals
\[
 \frac{23}{32}+\frac9{32}(1+u)^{-4}-(1+u)^{-9/8}\ge0
\]
by the weighted arithmetic--geometric mean inequality, with weights $23/32$ and $9/32$.

Averaging \eqref{eq:small-scalar} and using \eqref{eq:moments-three} and \eqref{eq:negative-moment}, with $\gamma=-2$, yields
\begin{equation}\label{eq:small-moment}
 \E\Phi(U)\le\frac{23}{64}\rho^2\left(\frac34+\frac{\rho^2}{4}\right)
 +\frac3{64}\left(2^{6\rho^2/(1-3\rho^2)}-1\right).
\end{equation}
For $v=\rho^2\le1/100$, use $e^x-1\le xe^x$ to bound the right side divided by $\rho^2$ by
\begin{equation}\label{eq:small-constant}
 \frac{23}{64}\left(\frac34+\frac1{400}\right)
 +\frac3{64}\frac{6\log2}{1-3/100}
 \exp\left(\frac{(6/100)\log2}{1-3/100}\right)
<\frac{49}{100}.\CASmark{04}
\end{equation}
Indeed, $\log2<7/10$ and $e^x\le(1-x)^{-1}$ for $0\le x<1$ bound the expression in \eqref{eq:small-constant} by $358267/742400<49/100$; by direct rational arithmetic. 
Since $\Psi(\rho)\ge\rho^2/2$, this proves \eqref{eq:Phi-target} strictly in the nonlocal case for $0<\rho\le1/10$.

\subsection{Three scalar majorants for the remaining correlations}\label{sec:majorants}
For $1/10\le\rho\le\rLow$, we use the parameters in \cref{tab:low}. Every terminating decimal in the table denotes the displayed rational number exactly.

\begin{table}[htbp]
\centering\small
\setlength{\tabcolsep}{3pt}
\caption{Scalar-majorant parameters at exponent $7/8$. \CASmark{05}}\label{tab:low}
\begin{tabular}{@{}ccccccc@{}}\toprule
$\rho$ interval &$a$ &$b$ &$\nu$ &$\eta$ &$C_{-1}$ &$C_{+1}$\\\midrule
$[.1,.26]$ &.73983025&.05001253&.04278380&.09539333&.00194881&.00260481\\
$[.26,.45]$&.83543237&.16917506&.00823695&.08958772&.01322350&.02614918\\
$[.45,.543]$&.86065349&.25388734&$-.08986605$&.07559427&.01118237&.08157943\\\bottomrule
\end{tabular}
\end{table}

\begin{lemma}[Global scalar majorants]\label{lem:scalar-majorants}
For every row of \cref{tab:low} and every $\varepsilon\in\{-1,1\}$,
\begin{equation}\label{eq:scalar-majorant}
 \Phi(u)\le au^2+(\nu-b\varepsilon)u+
 \eta\big((1+u)^{-1/2}-1\big)+C_\varepsilon,
 \qquad -1<u\le1.
\end{equation}
In addition, $a\rho^2\le b\le a\rho$ throughout the correlation interval of that row.
\end{lemma}

The proof of this lemma is given in Appendix~\ref{app:scalar-cert}. 

Substitute $(u,\varepsilon)=(U,g)$ in \eqref{eq:scalar-majorant}. Balance gives $\E U=0$ and $\Pp(g=1)=\Pp(g=-1)=1/2$. Using \eqref{eq:mixed} and \eqref{eq:negative-moment}, we obtain
\begin{equation}\label{eq:expected-majorant}
 \E\Phi(U)\le w(a\rho^2-b\rho)+\eta B(\rho)+\frac{C_{-1}+C_{+1}}2.
\end{equation}
The use of $\varepsilon=g$ is essential: it retains the mixed moment $\E[gU]$ until \eqref{eq:mixed} is applied.

Define, for the corresponding row of \cref{tab:low},
\begin{equation}\label{eq:Delta}
 \Delta(\rho)=\Psi(\rho)-w(a\rho^2-b\rho)-\eta B(\rho)
 -\frac{C_{-1}+C_{+1}}2.
\end{equation}

\begin{lemma}[Delta bound]\label{lem:Delta-bound}
We have
\begin{equation}\label{eq:Delta-bound}
 \Delta(\rho)>1/6000
 \quad\text{throughout each row's correlation interval}.
\end{equation}
\end{lemma}

The proof of this lemma is given in Appendix~\ref{app:expectation-interval}.

\begin{proposition}[Base exponent at low correlations]\label{prop:base-low}
At $\alpha=7/8$, inequality \eqref{eq:target} holds for every balanced Boolean indicator and every $0\le\rho\le\rLow$. At positive correlations in this interval, equality holds only for signed coordinate indicators.
\end{proposition}
\begin{proof}
If $m\ge23/50$, then $\delta\le27/100$, and the local comparison proved in \cref{sec:local} applies. Otherwise \cref{lem:W1} gives $W_1\le3/4$. For $0<\rho\le1/10$, use \eqref{eq:small-moment}--\eqref{eq:small-constant}. For $1/10\le\rho\le\rLow$, combine \eqref{eq:expected-majorant} and \eqref{eq:Delta-bound}. Both parts are strict in the nonlocal case. At $\rho=0$, $T_0f=1/2$ and equality is immediate.
\end{proof}

\section{Transfer from asymmetric Hellinger kernels}\label{sec:high}
We now assume \cref{ass:input} and prove the base exponent for $\rHigh\le\rho<1$. Fix a kernel from \cref{tab:source} whose band contains $s=\sqrt{1-\rho^2}$. Throughout this section $\alpha=7/8$ and
\begin{equation}\label{eq:H-target}
 H(t)=\frac{(1-t)^\alpha-(1-t)}{1-\alpha}.
\end{equation}
Thus $H(1-T_\rho f)$ is an affine normalization of the single-sided moment we seek to bound. Write
\begin{equation}\label{eq:AKM}
 A(t)=h(t),\qquad
 M(t)=\frac{A(t)}{1+A(t)\lambda(A(t))},\qquad
 K(t)=(2t-1)M(t),\qquad G_\pm(t)=A(t)\pm K(t).
\end{equation}
The functions $A,M$ are symmetric about $1/2$, whereas $K(1-t)=-K(t)$. In particular, $G_+=\G_\lambda$ and $G_-(t)=G_+(1-t)$.

\subsection{A global minorant with two double contacts}\label{sec:contact}

We seek the global minorant
\begin{equation}\label{eq:contact-minorant}
 H(t)\ge c_\rho A(t)+d_\rho K(t)+\ell_\rho t+e_\rho,
 \qquad c_\rho>d_\rho>0,
\end{equation}
such that the minorant is tangent to \(H\) at both \(p\) and \(\bar p\). Equivalently, the residual \(E\) (left side minus right side) satisfies
\[
E(p)=E'(p)=E(\bar p)=E'(\bar p)=0.
\]
These two contact points are exactly the two values of the noisy dictator. The strict inequality $c_\rho>d_\rho>0$ will allow us to combine the two source-kernel bounds with positive coefficients.

Let $\overline H'=(H'(p)+H'(\bar{p}))/2$ and define
\begin{align}
 c_\rho&=\frac{H'(p)-H'(\bar{p})}{2A'(p)},\label{eq:contact-c}\\
 d_\rho&=\frac{H(\bar{p})-H(p)-\rho\overline H'}{\rho^2M'(p)},\label{eq:contact-d}\\
 \ell_\rho&=\overline H'-d_\rho K'(p),\label{eq:contact-l}\\
 e_\rho&=\frac{H(p)+H(\bar{p})}2-c_\rho A(p)-\frac{\ell_\rho}{2}.\label{eq:contact-e}
\end{align}
The denominators will be shown positive below. Symmetry gives
\[
 A(\bar{p})=A(p),\quad A'(\bar{p})=-A'(p),\quad
 K(\bar{p})=-K(p),\quad K'(\bar{p})=K'(p).
\]
Since $\bar{p}-p=\rho$ and $M'(\bar{p})=-M'(p)$, by the third identity in \eqref{eq:AKM}, one also has
\begin{equation}\label{eq:contact-identity}
 K(\bar{p})-K(p)-\rho K'(p)=\rho^2M'(p).\CASmark{06}
\end{equation}
Equations \eqref{eq:contact-c} and \eqref{eq:contact-l} imply that  the residual $E$ satisfies $E'(p)=E'(\bar{p})=0$. Equation \eqref{eq:contact-d}, using \eqref{eq:contact-identity}, gives $E(\bar{p})-E(p)=0$, while \eqref{eq:contact-e} gives $E(p)+E(\bar{p})=0$. Thus
\begin{equation}\label{eq:contact-equalities}
 E(p)=E(\bar{p})=E'(p)=E'(\bar{p})=0.
\end{equation}
Rolle's theorem gives at least two distinct zeros of $E''$ inside $(p,\bar{p})$. If these are its only zeros on $(0,1)$ and the signs of $E''$ are $+,-,+$, integrating from the two double contacts shows $E\ge0$ on $[0,1]$, strictly away from $p,\bar{p}$. 
We next show this assumption. 

\subsection{The coefficient ratio}\label{sec:theta}
As $A'(p)>0$ and $H''<0$ (by definition), equation \eqref{eq:contact-c} shows that $c_\rho>0$. Recall $s=h(p)=A(p)$. Direct differentiation gives
\begin{equation}\label{eq:Mprime}
 M'(p)=A'(p)\frac{1-s^2\lambda'(s)}{(1+s\lambda(s))^2}.\CASmark{07}
\end{equation}
Define    $\theta=d_\rho/c_\rho$ and put
\begin{equation}\label{eq:z-parameter}
 z=(p/\bar{p})^{1/8},\qquad
 p=\frac{z^8}{1+z^8},\quad \bar{p}=\frac1{1+z^8},\quad
 \rho=\frac{1-z^8}{1+z^8},\quad s=\frac{2z^4}{1+z^8}.
\end{equation}
Substitution in \eqref{eq:contact-c}--\eqref{eq:contact-d} simplifies to
\begin{equation}\label{eq:theta}
 \theta=
 \frac{(1+s\lambda(s))^2}{1-s^2\lambda'(s)}
 \frac{\rho(1+z)+(16/7)(p-\bar p z)}{(1-z)\rho^2}.\CASmark{08}
\end{equation}
Indeed, by the definitions of $c_\rho$ and $d_\rho$,
\[
\theta=\frac{d_\rho}{c_\rho}
=\frac{A'(p)}{M'(p)}
\frac{2[H(\bar p)-H(p)]
-\rho[H'(p)+H'(\bar p)]}
{\rho^2[H'(p)-H'(\bar p)]}.
\]
Since $A(p)=s$ and
\[
M'(p)=A'(p)
\frac{1-s^2\lambda'(s)}
{(1+s\lambda(s))^2},
\]
the first factor equals
$(1+s\lambda(s))^2/(1-s^2\lambda'(s))$.
Moreover, writing $z=(p/\bar p)^{1/8}$,
direct differentiation of
$H(t)=8[(1-t)^{7/8}-(1-t)]$ gives
\[
H'(p)-H'(\bar p)=7p^{-1/8}(1-z)
\]
and
\[
\begin{aligned}
&2[H(\bar p)-H(p)]
-\rho[H'(p)+H'(\bar p)]\\
&\quad=7p^{-1/8}
\left[\rho(1+z)+\frac{16}{7}(p-\bar p z)\right].
\end{aligned}
\]
Substitution yields \eqref{eq:theta}.

\begin{lemma}[Positive mixing coefficients]\label{lem:theta}
For each of the three source bands, at every $0<s$ in that band,
\begin{equation}\label{eq:theta-bounds}
 1-s^2\lambda'(s)>0,\qquad 0<\theta<1.
\end{equation}
Consequently the coefficients in \eqref{eq:contact-minorant} satisfy $c_\rho>d_\rho>0$.
\end{lemma}

This lemma is proved in Appendix~\ref{app:theta-cert}. 

\subsection{Curvature ratios and unimodality}\label{sec:curvature}
Define
\begin{equation}\label{eq:RQ}
 R(t)=\frac{K''(t)}{A''(t)},\qquad
 Q(t)=\frac{H''(t)}{A''(t)}
 =2\alpha t^{3/2}(1-t)^{\alpha-1/2}
 =\frac74 t^{3/2}(1-t)^{3/8}.
\end{equation}
Here
\[
 A''(t)=-\frac1{2[t(1-t)]^{3/2}}<0.
\]
The role of these ratios is to reduce the sign changes of $E''$ to the shape of a single scalar function.

\begin{lemma}[Curvature shape]\label{lem:curvature}
For each kernel in \cref{tab:source},
\begin{equation}\label{eq:R-shape}
 R'(t)>0\quad(0<t<1),\qquad R(0+)=-1,\qquad R(1-)=1.
\end{equation}
For every fixed $0\le\vartheta<1$, the function
\begin{equation}\label{eq:unimodal-ratio}
 W_\vartheta(t)=\frac{Q(t)}{1+\vartheta R(t)}
\end{equation}
is strictly increasing up to one point and strictly decreasing thereafter. It tends to zero as $t\downarrow0$ and as $t\uparrow1$.
\end{lemma}
\begin{proof}
Write $x=2t-1$, $h=\sqrt{1-x^2}$ and
$M(h)=h/(1+h\lambda(h))$. Define
\[
 S(h)=(1+2h^2)M'(h)-(1-h^2)hM''(h),\qquad
 T(h)=S(h)-\frac{1-h^2}{h}S'(h).
\]
Since $A(t)=h$ and $K(t)=xM(h)$, we have
\[
x'=2,\qquad h'=-\frac{2x}{h},
\qquad A''(t)=h''=-\frac4{h^3}.
\]
Consequently,
\[
\begin{aligned}
K''(t)
&=4M'(h)h'
+x\bigl(M''(h)(h')^2+M'(h)h''\bigr)\\
&=-\frac{4x}{h^3}
\left[(1+2h^2)M'(h)
-(1-h^2)hM''(h)\right]\\
&=-\frac{4x}{h^3}S(h).
\end{aligned}
\]
Thus $R(t)=K''(t)/A''(t)=xS(h)$. Differentiating
once more and using $x^2=1-h^2$ gives
\[
R'(t)=2S(h)-\frac{2x^2}{h}S'(h)
=2\left[S(h)-\frac{1-h^2}{h}S'(h)\right]
=2T(h).
\]
The positive polynomial identities in Appendix~\ref{app:curvature-cert}
give $T>0$; also $S(0)=1$. This proves \eqref{eq:R-shape}.

It remains to locate the sign change of
\[
 V_\vartheta(t):=Q'(t)(1+\vartheta R(t))-
                         \vartheta Q(t)R'(t),
\]
which has the same sign as  $W_\vartheta'$.
On $0<t\le1/2$, set $y=\sqrt{t/(1-t)}$ and
\begin{equation}\label{eq:F01}
 f_0(y)=\frac{d}{dy}\log Q(t(y))
       =\frac{3(4-y^2)}{4y(1+y^2)},\qquad
 f_{1,j}(y)=(1+R_j(y))f_0(y)-\frac{dR_j(y)}{dy}.
\end{equation}
The sign of $V_\vartheta$ is the sign of
$(1-\vartheta)f_0+\vartheta f_{1,j}$.
Here $f_0>0$, and Appendix~\ref{app:unimodality-identities} proves
\begin{equation}\label{eq:F1-left}
 f_{1,j}(y)>0\qquad(0<y\le1).
\end{equation}
Thus $V_\vartheta>0$ on the left half of the interval.

On $1/2\le t\le4/5$, we have $0\le x\le3/5$, $h\ge4/5$,
and $R(t)\ge0$. The much weaker bound
\begin{equation}\label{eq:Tprime-bound}
 T_j'(h)<3\qquad(4/5\le h\le1),
\end{equation}
proved in Appendix~\ref{app:central-curvature}, suffices here: it gives
$R''(t)=-4xT_j'(h)/h\ge-12x/h$. Moreover
\[
 \frac{Q''(t)}{Q(t)}
 =-\frac{39+126x-105x^2}{16h^4}<0.
\]
Consequently, for $0\le\vartheta<1$,
\begin{equation}\label{eq:F-decreasing}
 \frac{V_\vartheta'(t)}{Q(t)}
 =\frac{Q''(t)}{Q(t)}(1+\vartheta R(t))-
       \vartheta R''(t)
 \le \frac{Q''(t)}{Q(t)}+\frac{12x}{h}<0.\CASmark{09}
\end{equation}
For the last inequality use $h^3\le h^2=1-x^2$ and the identity,
with $u=5x/3\in[0,1]$,
\begin{align*}
 39+126x-105x^2-192xh^3
 &\ge39-66x-105x^2+192x^3\\
 &=39(1-u)^3+\frac{387}{5}u(1-u)^2+
                          \frac{384}{125}u^3>0.
\end{align*}
Thus $V_\vartheta$ strictly decreases across this middle interval.

For $t\ge4/5$, $Q'\le0$ and $R'>0$, so
\begin{equation}\label{eq:F1-right}
 V_\vartheta(t)<0\quad(4/5\le t<1,\ 0<\vartheta<1).
\end{equation}
It follows that $V_\vartheta$ has exactly one zero when
$\vartheta>0$. This proves unimodality of $W_\vartheta$ for $ 0<\vartheta<1$.
For $\vartheta=0$, $Q'$ has its unique zero at $4/5$.
This proves unimodality for $ \vartheta=0$. Finally, $|R|<1$ and
$1+\vartheta R\ge1-\vartheta>0$; since $Q$ vanishes at both
endpoints, so does $W_\vartheta$.
\end{proof}

\begin{proposition}[Hellinger-to-power scalar minorant]\label{prop:minorant}
For every kernel and correlation in its source band with $0<\rho<1$, inequality \eqref{eq:contact-minorant} holds on $[0,1]$, with equality exactly at $p,\bar{p}$.
\end{proposition}
\begin{proof}
By \cref{lem:theta}, $0<\theta=d_\rho/c_\rho<1$. Using \eqref{eq:RQ}, the second derivative of the residual factors as
\begin{equation}\label{eq:E-factor}
 E''(t)=(-A''(t))(1+\theta R(t))
 \left(c_\rho-\frac{Q(t)}{1+\theta R(t)}\right).\CASmark{10}
\end{equation}
The first two factors are positive. By \cref{lem:curvature}, the ratio in the last factor is strictly unimodal, so the equation $E''=0$ has at most two solutions. The double contacts \eqref{eq:contact-equalities} force at least two distinct solutions in $(p,\bar{p})$. There are therefore exactly two, and the signs of $E''$ are $+,-,+$. Apply the two-contact sign observation to obtain the stated global inequality and its equality set.
\end{proof}

\subsection{Applying both orientations of the source estimate}\label{sec:high-expectation}
The curvature computation has an additional consequence:
\begin{equation}\label{eq:kernel-concavity}
 G_\pm''(t)=A''(t)(1\pm R(t))<0,\qquad 0<t<1.
\end{equation}
The kernels are continuous at the endpoints. By \eqref{eq:sorting}, \cref{ass:input} extends from increasing balanced indicators to all balanced indicators. Applying the extended estimate to $1-f$, and using the reflection identity for $G_\pm$, gives the two separate bounds
\begin{equation}\label{eq:both-source}
 \E G_+(T_\rho f)\ge s,\qquad
 \E G_-(T_\rho f)\ge s.
\end{equation}

\begin{proposition}[Base exponent at high correlations]\label{prop:base-high}
Under \cref{ass:input}, inequality \eqref{eq:target} at $\alpha=7/8$ holds for every balanced Boolean indicator and every $\rHigh\le\rho<1$. Equality forces $T_\rho f\in\{p,\bar{p}\}$ pointwise.
\end{proposition}
\begin{proof}
By positivity of the mixing coefficients,
\begin{equation}\label{eq:positive-combination}
 c_\rho A+d_\rho K
 =\frac{c_\rho+d_\rho}{2}G_+
 +\frac{c_\rho-d_\rho}{2}G_-.
\end{equation}
Apply \cref{prop:minorant} to $Z=1-T_\rho f$. The two bounds in \eqref{eq:both-source} remain valid for $Z$, with their orientations exchanged. Since $\E Z=1/2$, averaging the minorant gives
\begin{align*}
 \E H(Z)
 &\ge c_\rho s+\frac{\ell_\rho}{2}+e_\rho\\
 &=\frac{H(p)+H(\bar{p})}2,
\end{align*}
where $A(p)=s$ and \eqref{eq:contact-e} were used in the last step. Substituting \eqref{eq:H-target} proves \eqref{eq:target}. The source bands cover $0<s\le21/25$, which is exactly $\rHigh\le\rho<1$.

If equality holds in the moment comparison, it must hold in the pointwise minorant almost surely, because the coefficients in \eqref{eq:positive-combination} are positive and all preceding gaps are nonnegative. By \cref{prop:minorant}, $Z\in\{p,\bar{p}\}$, and the same is true of $T_\rho f$.
\end{proof}

\section{Completion, equality, and consequences}\label{sec:completion}

\begin{proof}[Proof of \cref{thm:main}]
By \cref{cor:up}, it suffices to consider the base exponent $7/8$. 
Combining \cref{prop:base-low,prop:base-high} implies the base exponent case. The endpoint $\rho=1$ is immediate from Booleanity. 

%

We conclude by characterizing the equality cases.
First, suppose that $0<\rho<1$ and $T_\rho f$
takes values only in $\{p,\bar p\}$.
Writing $g=2f-1$ and $U=T_\rho g$, we then have
$U\in\{-\rho,\rho\}$. Hence, by Parseval's identity,
\begin{equation}\label{eq:rigidity}
 \rho^2=\E U^2
 =\sum_{k\ge1}\rho^{2k}W_k(g),
 \qquad \sum_{k\ge1}W_k(g)=1.
\end{equation}
Since $\rho^{2k}<\rho^2$ for every $k\ge2$,
we must have $W_1(g)=1$. Thus
$g(x)=\sum_i a_i x_i$. As $g^2\equiv1$,
its degree-two Fourier coefficients
$2a_i a_j$ vanish for all $i\ne j$.
Together with $\sum_i a_i^2=1$, this forces
$g=\pm x_i$ for some $i$.

We now apply this rigidity argument to the equality
cases. At the base exponent,
\cref{prop:base-low} is strict for non-dictators
in the low-correlation range, while equality in
\cref{prop:base-high} forces
$T_\rho f\in\{p,\bar p\}$ in the remaining range.
For larger exponents, \cref{cor:up} reduces equality
to the same two-point condition. In either case,
\eqref{eq:rigidity} shows that $f$ must be a
signed dictator. Conversely, every signed
dictator attains equality by
\eqref{eq:dictator-benchmark}.

The Shannon inequality follows by subtracting
$1/2$ from both sides of \eqref{eq:main},
dividing by $\alpha-1<0$, and letting
$\alpha\uparrow1$. Since the cube is finite,
the limit passes through the expectation.
Finally, at $\rho=0$ and $\rho=1$, the identities
$T_0f=1/2$ and $T_1f=f$ show that every
balanced indicator attains equality.
\end{proof}

\section*{Acknowledgements}

\emph{Generative-AI use disclosure:}
ChatGPT (GPT-6 Astra) was used to explore proof strategies,
develop and refine mathematical arguments and exposition,
and assist in drafting the manuscript.
The proof was developed during the author's efforts to resolve
the symmetric version of Conjecture~\ref{conj:main}, studied
in \cite{YuSymmetric}. An earlier version relied on both
\cite{OpenAI} and \cite{YuSymmetric} and involved extensive
numerical verification. In particular, the high-correlation
regime was divided into two subintervals, treated using
the square-root kernel and the Hellinger kernel, respectively.
Through iterative exchanges with ChatGPT, the proof was
substantially simplified. The dependence on \cite{YuSymmetric}
was eliminated, and the numerical verification procedures
were replaced by analytic arguments, supplemented by exact
symbolic computations for certain lengthy algebraic identities
and coefficient inequalities. The resulting proof builds upon
\cite{OpenAI} and \cite{Yu}.
The author assumes full responsibility for the correctness
and content of the final manuscript.

\emph{Funding:} This work was supported by the National Key Research
and Development Program of China under grant 2023YFA1009604 and the
NSFC under grant 62101286.

\appendix
\numberwithin{equation}{section}
\numberwithin{table}{section}
\setcounter{equation}{0}\setcounter{table}{0}
\section*{Appendices}



\section{Proof of Lemma \ref{lem:local-derivatives}}
\label{app:local-interval}

\paragraph{Analyticity and a uniform complex bound.}
Recall $a=7/8$, $d=27/100$, $m=23/50$, and $\ell=355/512$.
Put $R=9/10$ and $r=R^2=81/100$.
The logarithm of $1+z$ is analytic on $|z|\le R$, with
$|\log(1+z)|\le\log10$.
For
\[
 A(z)=2d e^{-\ell z^2/8}+m(1+z)^{1-z^2/8},
\]
the identity $2d+m=1$,  the triangle
inequality, and $|e^w-1|\le e^{|w|}-1$ give 
\begin{align}\label{eq:local-complex-disk}
|A(z)-1|
={}&\Big|2d(e^{-\ell z^2/8}-1)+mz \notag\\
&\quad +m(1+z)
\big(e^{-z^2\log(1+z)/8}-1\big)\Big|\notag\\
\le{}&
2d(e^{\ell R^2/8}-1)+mR\notag\\
&+m(1+R)
(e^{R^2\log 10/8}-1)\notag\\
<{}&
\frac{54}{100}\frac{74}{1000}
+\frac{46}{100}
\left(\frac9{10}+\frac{19}{10}
\frac{268}{1000}\right)\notag\\
={}&\frac{688192}{10^6}<\frac7{10}\CASmark{11}.
\end{align}
For an elementary rational check of the the second  inequality,
$\ell<7/10$ and $\log10<7/3$ give
$\ell R^2/8<71/1000$ and $R^2\log10/8<237/1000$.
The series bound
\[
 e^x\le1+x+\frac{x^2}{2(1-x/3)}\qquad(0\le x<3)
\]
then gives $e^{71/1000}<1074/1000$ and
$e^{237/1000}<1268/1000$, which prove
\eqref{eq:local-complex-disk} by rational arithmetic.

Thus the analytic logarithm of $A$ satisfies
$|\log A|\le\log(10/3)$. Moreover
\[
 \left|\frac{a}{1-z^2/8}\right|\le\frac{700}{719}<1.
\]
The first term $A^{\frac{a}{1-z^2/8}}$ in $\mathcal J$ (defined in \eqref{eq:local-analytic-branch}) consequently has modulus  $|A^{\frac{a}{1-z^2/8}}|\le e^{|\frac{a}{1-z^2/8}\log A|} \le 10/3$, while $|(1+z)^a|\le(19/10)^a<2$.
It follows that $|\mathcal J(z)|<6$ for $|z|\le R$.
Its even part defines the analytic function
\[
 \mathcal F(v)=\sum_{k\ge1}d_kv^k,\qquad |v|\le r,
 \qquad |d_k|\le6r^{-k},
\]
where the last assertion follows since 
 $\mathcal J$ is analytic on $|z|\le R$,
$|\mathcal J(z)|<6$, and thus, Cauchy's coefficient estimate gives
\[
|d_k|=|[z^{2k}]\mathcal J(z)|
\le 6R^{-2k}=6r^{-k},
\qquad r=R^2.
\]

\paragraph{Bounds on Taylor expansion.}

Since $\ell\in\mathbb Q$ and $A(0)=2d+m=1$,
we have $A(z)\in1+z\mathbb Q[[z]]$. Consequently,
\[
\log A(z)\in z\mathbb Q[[z]],
\qquad
\exp\left(\frac{7}{8-z^2}\log A(z)\right)
\in1+z\mathbb Q[[z]].
\]
Thus every Taylor coefficient $d_k$ is rational.

The exact algebraic coefficients through $k=14$ are specified in
Appendix~\ref{app:taylor-data}. In particular,
\begin{equation}\label{eq:local-coefficient-bounds}
 d_2=-\frac{24329604577383}{3355443200000000}<-\frac7{1000},
 \qquad 0<d_k<\frac1{1000}\quad(3\le k\le14).\CASmark{12}
\end{equation}

For $0\le v\le\rLow^2<3/10$ and $q=v/r<3/8$,
\begin{align*}
 \left|\sum_{k\ge15}k(k-1)d_kv^{k-2}\right|
 &\le\frac6{r^2}\sum_{k\ge15}k(k-1)q^{k-2} \\
 &=\frac6{r^2} q^{13}\sum_{j\ge0} \bigl(210+30j+j(j-1)\bigr)q^j\\
 &= \frac6{r^2}q^{13}
 \left(\frac{210}{1-q}+\frac{30q}{(1-q)^2}
                       +\frac{2q^2}{(1-q)^3}\right)\\
 &\le\frac{10421055}{1073741824}<\frac1{100}\CASmark{13}.
\end{align*}
Likewise, by \eqref{eq:local-coefficient-bounds},
\[
 \sum_{k=3}^{14}k(k-1)d_kv^{k-2} \le  \frac1{1000}  \sum_{k=3}^{\infty}k(k-1)v^{k-2}
=\frac1{1000}\left(\frac2{(1-v)^3}-2\right)
 <\frac1{250}.
\]
Combining these estimates gives
\[
 \mathcal F''(v)<-\frac{14}{1000}+\frac1{250}+\frac1{100}=0.
\]
This proves concavity on the whole required interval analytically.

Finally, substitution of the single rational value
$v=\rLow^2=294849/10^6$ gives
\begin{equation}\label{eq:local-single-endpoint}
 \frac{2^{-7/8}\mathcal F(294849/10^6)}{294849/10^6}
 >\frac{4177}{10^8}>\frac1{25000}.\CASmark{14}
\end{equation}
The value is approximately $0.00004177668374$. We complete the proof of Lemma  \ref{lem:local-derivatives}.

\subsection{Exact Taylor coefficients used above}
\label{app:taylor-data}
This subsection supplies the finite algebraic data in
\eqref{eq:local-coefficient-bounds}.
All coefficients can be reconstructed by the following
explicit formal-series identities, truncated after degree $28$:
\begin{align*}
 E(z)&=\exp\!\left(-\frac{z^2}{8}\log(1+z)\right),\\
 A(z)&=\frac{27}{50}\exp(-355z^2/4096)
              +\frac{23}{50}(1+z)E(z),\\
 B(z)&=\exp\!\left(\frac{7}{8-z^2}\log A(z)\right),\\
 d_k&=[z^{2k}]B(z)-\binom{7/8}{2k}.
\end{align*}
For clarity, if $C(z)=\sum c_jz^j$ and $c_0=0$, the coefficients
of $e^{C(z)}=\sum e_jz^j$ are determined solely by
\[
 e_0=1,\qquad e_n=\frac1n\sum_{j=1}^n j c_j e_{n-j}.
\]
If $ A(z)=\sum a_jz^j$, $A(0)=1$, and $\log A(z)=\sum l_jz^j$, use
$A'=A(\log A)'$ to obtain
\[
 l_0=0,\qquad
 l_n=a_n-\frac1n\sum_{j=1}^{n-1}j l_j a_{n-j}.
\]
These are polynomial coefficient identities over $\mathbb Q$;
there is no rounding or interval decision.

The following table gives the exact fractions $d_k=N_k/D_k$.
To keep long integers legible, let
$\langle a_0,\ldots,a_m\rangle_{10^9}
=\sum_{j=0}^m a_j10^{9(m-j)}$; entries after the first are
base-$10^9$ digits, including possible leading zeros.
The bounds in \eqref{eq:local-coefficient-bounds} follow directly
by multiplying these positive denominators.
{\footnotesize\setlength{\tabcolsep}{3pt}
\begin{longtable}{@{}cp{.59\textwidth}p{.32\textwidth}@{}}\toprule
$k$ & $N_k$ (base-$10^9$ notation) & $D_k$\\\midrule\endfirsthead
\toprule $k$ & $N_k$ & $D_k$\\\midrule\endhead\bottomrule\endfoot
1 & $\begin{gathered}\langle 88641\rangle_{10^9}\end{gathered}$ & $2^{16}\,5^{4}$\\[4pt]
2 & $\begin{gathered}-\langle 24329,604577383\rangle_{10^9}\end{gathered}$ & $2^{33}\,5^{8}$\\[4pt]
3 & $\begin{gathered}\langle 9,485077841,592463719\rangle_{10^9}\end{gathered}$ & $2^{49}\,5^{11}$\\[4pt]
4 & $\begin{gathered}\langle 2,735295087,887220546,594654111\rangle_{10^9}\end{gathered}$ & $2^{67}\,5^{15}$\\[4pt]
5 & $\begin{gathered}\langle 90547576,814225909,624259534,850997043\rangle_{10^9}\end{gathered}$ & $2^{83}\,5^{19}$\\[4pt]
6 & $\begin{gathered}\langle 5898065,861502085,686308818,713157094,\\930678601\rangle_{10^9}\end{gathered}$ & $2^{100}\,5^{23}$\\[4pt]
7 & $\begin{gathered}\langle 2138942,965422397,987495840,865803429,\\529986916,919553941\rangle_{10^9}\end{gathered}$ & $2^{116}\,5^{27}\,11$\\[4pt]
8 & $\begin{gathered}\langle 2037506,211707916,756816311,917023454,\\986548560,990575052,766437153\rangle_{10^9}\end{gathered}$ & $2^{135}\,3\,5^{31}\,13$\\[4pt]
9 & $\begin{gathered}\langle 5356704,930075102,873971444,942403012,\\669960049,567259968,649959848,742540741\rangle_{10^9}\end{gathered}$ & $2^{151}\,3\,5^{35}\,7\,11\,13$\\[4pt]
10 & $\begin{gathered}\langle 18927836,999824278,239893022,489002987,\\232311039,361707706,364577211,808196064,\\542580539\rangle_{10^9}\end{gathered}$ & $2^{168}\,3^{2}\,5^{39}\,7\,11\,13\,17$\\[4pt]
11 & $\begin{gathered}\langle 1401880,459171512,631486290,620640917,\\519066199,394189313,531332056,563520037,\\361985486,972817659\rangle_{10^9}\end{gathered}$ & $2^{184}\,5^{43}\,7\,11\,13\,17\,19$\\[4pt]
12 & $\begin{gathered}\langle 1790263,175021975,597155075,419573113,\\602782111,758872200,802180824,520393010,\\970880964,118287260,953984569\rangle_{10^9}\end{gathered}$ & $2^{202}\,3^{2}\,5^{47}\,7\,11\,13\,17\,19$\\[4pt]
13 & $\begin{gathered}\langle 3526950,35938785,965743790,262872311,\\541565857,308134246,714664543,788340804,\\305820340,650452986,18774397,700863131\rangle_{10^9}\end{gathered}$ & $2^{218}\,3^{2}\,5^{52}\,7\,11^{2}\,13\,17\,19$\\[4pt]
14 & $\begin{gathered}\langle 28618,367266670,796907149,387227216,\\735716086,481177967,276851253,509632921,\\757956226,646347429,764465436,355616418,\\319238731\rangle_{10^9}\end{gathered}$ & $2^{235}\,3^{3}\,5^{56}\,7^{2}\,11\,13\,19$\\[4pt]
\end{longtable}}

\section{Proof of Lemma \ref{lem:W1}}
\label{app:W1-proof}

We first prove an absolute-moment estimate for a Rademacher sum
\begin{align}\label{eq:appendix-rademacher}
 Z&=\sum_i b_iX_i,\quad \sum_i b_i^2=1,
 \quad \max_i b_i^2\le c:=\frac{17}{60}
\quad \implies \quad \E|Z|<\frac{433}{500}<\sqrt{\frac34}.
\end{align}
Consider the degree-five polynomial
\begin{equation}\label{eq:W1-Q}
 Q(t)=\frac{3809025+5707611t-726070t^2+58998t^3
                -2235t^4+31t^5}{8847360}.
\end{equation}
For every $u\ge0$, the exact factorization \CASmark{15}
\begin{align}\label{eq:W1-factor}
 8847360\bigl(Q(u^2)-u\bigr)
 ={}&(u-1)^2(u-3)^2(u-5)^2\\[-2pt]
 &\times\bigl(31u^4+558u^3+3872u^2
                  +12594u+16929\bigr)\notag
\end{align}
shows that $Q(u^2)\ge u$.  Consequently,
\begin{equation}\label{eq:W1-majorant}
 P(t):=\frac35 Q\!\left(\frac{25t}{9}\right)
 \ge\sqrt t,\qquad t\ge0.
\end{equation}
Write $P(t)=\sum_{j=0}^5c_jt^j$.

Put
\[
 x=\sum_i b_i^4,\quad y=\sum_i b_i^6,\quad
 z=\sum_i b_i^8,\quad v=\sum_i b_i^{10}.
\]
Cauchy--Schwarz, $ \sum_i b_i^2=1$, and $b_i^2\le c$ imply
\begin{equation}\label{eq:W1-moment-constraints}
 0<x\le c,\qquad x^2\le y\le cx,\qquad
 z\ge\frac{y^2}{x},\qquad v\le cz.
\end{equation}
The requisite Rademacher moment identities are \CASmark{16}
\begin{align*}
 \E Z^2&=1,\\
 \E Z^4&=3-2x,\\
 \E Z^6&=15-30x+16y,\\
 \E Z^8&=105-420x+140x^2+448y-272z,\\
 \E Z^{10}&=945-6300x+6300x^2+10080y
              -6720xy-12240z+7936v.
\end{align*}
These moment identities follow from the moment-generating function of \(Z\). Since the Rademacher random variables \(\varepsilon_i\) are independent, we have
\[
\mathbb E e^{tZ}=\prod_i\cosh(a_it),
\]
and hence
\[
\log\mathbb E e^{tZ}
=\frac{t^2}{2}-\frac{xt^4}{12}+\frac{yt^6}{45}
-\frac{17zt^8}{2520}+\frac{31vt^{10}}{14175}
+O(t^{12}),
\]
and
\begin{align*}
\mathbb E e^{tZ}
={}&1+\frac{t^2}{2!}
+\frac{3-2x}{4!}t^4+\frac{15-30x+16y}{6!}t^6 +\frac{105-420x+140x^2+448y-272z}{8!}t^8\\
&+\frac{945-6300x+6300x^2+10080y-6720xy-12240z+7936v}{10!}t^{10} +O(t^{12}),
\end{align*}
where we used \(\sum_i a_i^2=1\) and the Taylor expansion of \(\log\cosh t\). 
Comparing the coefficients of \(t^{2k}\) with those in
\[
\mathbb E e^{tZ}
=\sum_{k=0}^{\infty}\frac{\mathbb E Z^{2k}}{(2k)!}t^{2k}
\]
yields the stated moment formulas.

Substituting these moment identities into $\E P(Z^2)$, first using
$v\le cz$, makes the coefficient of $z$ equal to
\[
 -\frac{1664140625}{1632586752}<0.
\]
We may therefore also use $z\ge y^2/x$.  With $k=y/x$,
\eqref{eq:W1-moment-constraints} gives $0<x\le k\le c$.
The resulting exact polynomial expansion  is
\begin{equation}\label{eq:W1-one-variable-reduction}
 \E P(Z^2)
 \le \left(1-\frac{x}{k}\right)C_0
       +\frac{x}{k}F(k)-x(k-x)J(k),\CASmark{17}
\end{equation}
where
\begin{align*}
 C_0&=\frac{34711015}{40310784},\\
 F(k)&=\frac{5}{3265173504}
  \bigl(562318443-139168800k+1150000875k^2
            -2191437500k^3\bigr),\\
 J(k)&=\frac{112109375}{120932352}
             -\frac{423828125}{181398528}k.
\end{align*}
Set
\[
 C=\frac{433}{500},\qquad r=\frac{27}{100},\qquad
 \zeta=\frac{1899475}{483729408},\qquad
 \eta=\frac{27774371}{362797056000}.
\]
The following single identity \CASmark{18} exhibits all the slack in
\eqref{eq:W1-one-variable-reduction}:
\begin{align}\label{eq:W1-tangent}
 &C-\left[\left(1-\frac{x}{k}\right)C_0+
             \frac{x}{k}F(k)-x(k-x)J(k)\right]\notag\\
 &\quad=\left(1-\frac{x}{k}\right)(C-C_0)+x(k-x)J(k)\notag\\
 &\qquad\quad+\frac{x}{k}\left[
     \eta+\zeta(c-k)+(k-r)^2\left(
       \frac{6180625}{120932352}
       +\frac{2739296875}{816293376}k\right)\right],
\end{align}
where the expression inside the last square bracket is just $C-F(k)$.
Since $F$ is a cubic polynomial, its Taylor expansion
at $r$ is exact:
\[
\begin{aligned}
F(k)
={}&F(r)+\zeta(k-r)-(k-r)^2(A+Bk),
\end{aligned}
\]
where
\[
\zeta=F'(r),\qquad
A=-\frac{F''(r)}2+\frac{rF'''(r)}6,
\qquad B=-\frac{F'''(r)}6.
\]
Direct substitution of $r=27/100$ gives
\[
A=\frac{6180625}{120932352}>0,\qquad
B=\frac{2739296875}{816293376}>0.
\]
Defining
\[
\eta=C-F(r)-\zeta(c-r)>0,
\]
we obtain
\[
C-F(k)
=\eta+\zeta(c-k)+(k-r)^2(A+Bk),
\]
which yields the claimed decomposition.

Every summand on the right of \eqref{eq:W1-tangent}
 is nonnegative for $0<x\le k\le c$,
and the term $(x/k)\eta$ is positive. Indeed,
\[
 C-C_0=\frac{24765493}{5038848000}>0,\qquad
 J(k)\ge J(c)=\frac{576953125}{2176782336}>0.
\]
Together with \eqref{eq:W1-majorant} and
\eqref{eq:W1-one-variable-reduction}, this proves $\E|Z|<C<\sqrt{3/4}$,
proving
\eqref{eq:appendix-rademacher}.

We now deduce \cref{lem:W1}.  Set $a_i=\widehat g(\{i\})$,
$W=\sum_i a_i^2$ and $L=\sum_i a_iX_i$.  Booleanity gives
\[
 W=\E[gL]\le\E|L|.
\]
If $W>3/4$, let $b_i=a_i/\sqrt W$.  Then $\sum_i b_i^2=1$ and
\[
 \max_i b_i^2\le\frac{(23/50)^2}{W}
 <\frac{529}{1875}<\frac{17}{60}.
\]
Applying \eqref{eq:appendix-rademacher} to $Z=L/\sqrt W$ yields
\[
 \sqrt W\le\E|Z|<\frac{433}{500}<\sqrt{\frac34},
\]
contradicting $W>3/4$.  This proves \cref{lem:W1}.


\section{Proof of  Lemma \ref{lem:scalar-majorants}}
\label{app:polynomial}\label{app:scalar-cert}
For each row of \cref{tab:low}, put
\begin{align}\label{eq:residual-t}
 \mathcal R_{i,\varepsilon}(t)={}&
 a_i(t-1)^2+(\nu_i-b_i\varepsilon)(t-1)+C_{i,\varepsilon}
 +\eta_i(t^{-1/2}-1)\notag\\
 &-\frac87-8t+\frac{64}{7}t^{7/8},\qquad t>0.
\end{align}
We prove its positivity on the entire positive half-line, not merely
at sampled points.

\subsection{The possible minima lie in two convex regions}
Direct differentiation gives \CASmark{19}
\begin{align}\label{eq:majorant-derivatives}
 \mathcal R'_{i,\varepsilon}(t)
 &=2a_i(t-1)+\nu_i-b_i\varepsilon
      -\frac{\eta_i}{2}t^{-3/2}-8+8t^{-1/8},\notag\\
 \mathcal R''_{i,\varepsilon}(t)
 &=2a_i+\frac{3\eta_i}{4}t^{-5/2}-t^{-9/8},\notag\\
 \mathcal R'''_{i,\varepsilon}(t)
 &=\frac38t^{-7/2}(3t^{11/8}-5\eta_i).
\end{align}
Thus $\mathcal R''$ strictly decreases up to
$\tau_i=(5\eta_i/3)^{8/11}$ and strictly increases afterwards.
Let $l=13/100$ and $h=13/20$. The following fixed-value
comparisons follow by substitution of the three rows:
\begin{equation}\label{eq:majorant-shape-points}
\begin{gathered}
 l<\tau_i<h,\qquad
 \mathcal R''_i(l)>1,\quad
 \mathcal R''_i(1/4)<0,\quad
 \mathcal R''_i(h)>1/20,\\
 \mathcal R'_{i,-1}(l)>0,\qquad
 \mathcal R'_{i,-1}(h)<0.
\end{gathered}\CASmark{20}
\end{equation}
For clarity, loose bounds sufficient for these comparisons are
\[
\begin{array}{c|rrrrr}
 i&\mathcal R''_i(l)&\mathcal R''_i(1/4)&\mathcal R''_i(h)
 &\mathcal R'_{i,-1}(l)&\mathcal R'_{i,-1}(h)\\\hline
1&>3&<-9/10&>3/50&>1/10&<-7/100\\
2&>5/2&<-9/10&>6/25&>9/100&<-1/20\\
3&>1&<-6/5&>13/50&>9/50&<-3/50
\end{array}
\]
All entries concern individual algebraic expressions. They can, for
example, be checked by rational bounds for the positive eighth root,
followed by eighth powers; they are not interval-wide sign tests.

It follows that $\mathcal R''_i$ has exactly two positive zeros,
both between $l$ and $h$, and has signs $+,-,+$. Together with
\eqref{eq:majorant-shape-points}, this implies that
$\mathcal R_{i,-1}$ has no local minimum in $(l,h)$. 
On the two outer regions we have the uniform analytic bounds
\begin{equation}\label{eq:majorant-convex-regions}
 \mathcal R''_i\ge1\quad(0<t\le l),\qquad
 \mathcal R''_i\ge1/20\quad(t\ge h).
\end{equation}


\subsection{Quadratic lower bounds at nine fixed points}
If for a function $F$, $F''\ge\kappa>0$ on an interval and $t_0$ belongs to it,
integration twice, followed by completing the square, gives
\begin{equation}\label{eq:convex-minimum-bound}
 F(t)\ge F(t_0)+F'(t_0)(t-t_0)+\frac\kappa2(t-t_0)^2
 \ge F(t_0)-\frac{F'(t_0)^2}{2\kappa}.
\end{equation}
Use the following rational choices of $t_0$:
\begin{table}[htbp]\centering
\caption{Evaluation points in the analytic proof of
\cref{lem:scalar-majorants}. Every decimal is an exact rational. \CASmark{21}}
\label{tab:scalar-points}
\begin{tabular}{@{}cccc@{}}\toprule
Row & $\varepsilon=-1$, $t_0<l$ & $\varepsilon=-1$, $t_0>h$
    & $\varepsilon=1$, $t_0>h$\\\midrule
1&$0.10862$&$0.91098$&$1.09268$\\
2&$0.10933$&$0.78595$&$1.24498$\\
3&$0.08928$&$0.81246$&$1.40710$\\\bottomrule
\end{tabular}
\end{table}
At each of these nine points, direct evaluation gives the deliberately
loose strict bounds
\begin{equation}\label{eq:scalar-point-values}
 \mathcal R_{i,\varepsilon}(t_0)>\frac{3}{2\cdot10^6},\qquad
 |\mathcal R'_{i,\varepsilon}(t_0)|<\frac1{10000}.\CASmark{21}
\end{equation}
These are isolated algebraic-number comparisons: if
$y=t_0^{1/8}$, both expressions in
\eqref{eq:scalar-point-values} are rational functions of $y$, and
rational bounds on $y$ are verified simply by raising them to the
eighth power. For reference the smallest residual in the table is
larger than $1.6839\cdot10^{-6}$, well above the displayed lower bound.

Equations \eqref{eq:convex-minimum-bound}--\eqref{eq:scalar-point-values}
show that each relevant residual is larger than
$3/(2\cdot10^6)-10^{-7}>0$ for $0<t \le l$ and for $t\ge h$ when $\varepsilon=-1$, and for $t\ge h$ when  $\varepsilon=1$.
The absence of an interior minimum propagates positivity of
$\mathcal R_{i,-1}$ across $[l,h]$. Finally,
\[
 \mathcal R_{i,1}(t)-\mathcal R_{i,-1}(t)
 =C_{i,1}-C_{i,-1}+2b_i(1-t)>0\qquad(0<t\le h),
\]
since $C_{i,1}>C_{i,-1}$ and $h<1$.
This establishes
\cref{lem:scalar-majorants}.

\section{Proof of Lemma \ref{lem:Delta-bound}}
\label{app:expectation-interval}
Differentiation shows that $\Psi''$ is increasing for $\rho\ge0$.
Writing $L=\log2$ and
\[
G(\rho)=\frac{3L}{4}
\frac{\rho^2}{1-\frac32\rho^2},
\qquad B(\rho)=e^{G(\rho)}-1,
\]
we have
\[
G''(\rho)=\frac{3L}{2}
\frac{1+\frac92\rho^2}
{(1-\frac32\rho^2)^3}>0.
\]
Thus,
\[
B''(\rho)
=e^{G(\rho)}\bigl(G''(\rho)+G'(\rho)^2\bigr)
\ge G''(\rho).
\]
Moreover, for $0\le x<1$,
\[
\frac{1+3x}{(1-x)^3}
=\sum_{n=0}^\infty(n+1)(2n+1)x^n
\ge1+6x+15x^2.
\]
Taking $x=3\rho^2/2$ yields
\begin{equation}\label{eq:Bpp-analytic}
B''(\rho)\ge\frac{3L}{2}
\left(1+9\rho^2+\frac{135}{4}\rho^4\right).\CASmark{22}
\end{equation}
In particular, $B''>1$ for all three bands, and $B''>4$ for
$\rho\ge9/20$, using $L>69/100$.
At the three right endpoints, the single-value comparisons
\[
 \Psi''(13/50)<109/100,\quad
 \Psi''(9/20)<131/100,\quad
 \Psi''(543/1000)<152/100\CASmark{23}
\]
are immediate from \eqref{eq:Phi-curvature}.
Since $w=3/4$, we obtain on the respective intervals,
\[
\begin{array}{c|c}
 i&\text{upper bound for }\Delta_i''=\Psi''-\tfrac32a_i-\eta_iB''\\\hline
1&109/100-3a_1/2-\eta_1<-11/100\\
2&131/100-3a_2/2-\eta_2<-3/100\\
3&152/100-3a_3/2-4\eta_3<-7/100
\end{array}
\]
These derivative signs follow on the full bands from monotonicity.

The only remaining comparisons are the six fixed endpoint values
of $\Delta_i$ as explicitly defined in \eqref{eq:Delta}:
\begin{table}[htbp]\centering
\caption{Single-expression lower bounds for the expectation gaps.
All displayed integers are divided by $10^9$. \CASmark{24}}
\label{tab:endpoint-gaps}
\begin{tabular}{@{}ccc@{}}\toprule
Row & Left endpoint & Right endpoint\\\midrule
1&430611&430612\\
2&1642292&1642293\\
3&1991781&187631\\\bottomrule
\end{tabular}
\end{table}
Each bound exceeds $1/6000$. They may be obtained by the usual
Taylor series for the logarithm and exponential with remainder,
or direct fixed-point evaluation. Concavity gives
$\Delta_i(\rho)\ge\min\{\Delta_i(\ell_i),\Delta_i(h_i)\}>1/6000$
on each whole band. This proves \eqref{eq:Delta-bound}.

\section{Proof of Lemma  \ref{lem:theta}}
\label{app:theta-cert}
Write $\lambda_j(s)=L_j+U_js+V_js^2$.
Since $U_j+2V_j\le4/25$, on $[0,1]$ we have
\[
 1-s^2\lambda'_j(s)\ge1-\max(0,U_j+2V_j)\ge\frac{21}{25}\CASmark{25}.
\]
This proves $M'(p)>0$.

Note that $H'$ is strictly concave, since
\[
 H'''(t)=\alpha(\alpha-2)(1-t)^{\alpha-3}<0.
\]
The Hermite--Hadamard inequality for this concave function gives
\[
 H(\bar{p})-H(p)=\int_p^{\bar{p}} H'(t)\,dt>\rho\overline H'.
\]
Hence $d_\rho>0$, and also $\theta>0$ since $c_\rho>0$. 


\paragraph{A bound independent of the source kernel.}
Let $z=(p/\bar p)^{1/8}$, $w=z+z^{-1}>2$, $v=w-2$, and put
\[
 \mathcal H=w^4-4w^2+2,\qquad
 \mathcal J=7w^3+12w^2-6w-8,\qquad
 \mathcal K=7(w+2)w^2(w^2-2)^2.
\]
%
Consequently,
\[
s= \frac2{z^4+z^{-4}}
=\frac2{\mathcal H}.
\]
Recall 
\[
p=\frac{z^8}{1+z^8},\qquad
\bar p=\frac1{1+z^8},\qquad
\rho=\frac{1-z^8}{1+z^8}.
\]
Let $\Xi(\rho)$ denote the second factor in \eqref{eq:theta}. Substituting these expressions above gives
\[
\Xi=
\frac{(1+z^8)
(7z^6+12z^5+15z^4+16z^3
+15z^2+12z+7)}
{7(1+z)^2(1+z^2)^2(1+z^4)^2}.
\]
Using $w=z+z^{-1}$, the numerator equals
$z^7\mathcal H\mathcal J$, while the
denominator equals $z^7\mathcal K$.
Hence 
\[
\Xi=\mathcal H\mathcal J/\mathcal K\CASmark{26}.
\]

All displayed denominators are positive. The exact factorization \CASmark{27}
\begin{align}\label{eq:universal-ratio-factorization}
 &6\mathcal H(\mathcal K-\mathcal H\mathcal J)
            -23\mathcal H\mathcal J+6\mathcal J\notag\\
 &\quad=v^2\bigl(29808+103592v+148396v^2+115322v^3
             +53774v^4\notag\\
 &\hspace{36mm}{}+15571v^5+2760v^6+276v^7+12v^8\bigr)>0
\end{align}
is equivalent, after multiplying by positive denominators, to
\begin{equation}\label{eq:universal-Xi-bound}
 0<\Xi(\rho)<\frac1{1+23s/12-s^2/4},\qquad 0<\rho<1.\CASmark{28}
\end{equation}
It applies at every $0<\rho<1$, independently of $j$.

\paragraph{Checking the remaining quintics by their shape.}
Define $E_j$ by  
\begin{equation}\label{eq:mixing-quintic}
 E_j(s)=\frac{(1+23s/12-s^2/4)(1-s^2\lambda'_j(s))
                      -(1+s\lambda_j(s))^2}{s}.
\end{equation}
Its coefficients  display together:
\begin{equation}\label{eq:quintic-data}
 E_j(s)=\frac1{d_j}\sum_{k=0}^{5} e_{jk}s^k,\CASmark{29}
\end{equation}
\begin{center}\small\setlength{\tabcolsep}{4pt}
\begin{tabular}{@{}crrrrrrr@{}}\toprule
$j$&$d_j$&$e_{j0}$&$e_{j1}$&$e_{j2}$&$e_{j3}$&$e_{j4}$&$e_{j5}$\\\midrule
1&7500&275&1398&3529&$-15132$&3105&$-675$\\
2&750000&108500&119253&351067&$-1244428$&243000&$-46875$\\
3&3000000&698000&138108&1458670&$-3567285$&631800&$-97200$\\\bottomrule
\end{tabular}
\end{center}
In all three rows, $E_j'(0)>0$ and
\[
 d_jE_j^{(3)}(s)=6e_{j3}+24e_{j4}s+60e_{j5}s^2
 \le6e_{j3}+24e_{j4}<0\qquad(0\le s\le1).
\]
Thus $E_j'$ is strictly concave. Since it is positive at zero, it
can change sign at most once, from positive to negative.
Consequently $E_j$ has no interior minimum on $[0,b_j]$,
where $(b_1,b_2,b_3)=(4/25,3/10,21/25)$.
The two endpoint comparisons are
\[
\begin{array}{c|cc}
 j&E_j(0)&E_j(b_j)\\\hline
 1&11/300&68880969/976562500>7/100\\
 2&217/1500&576507071/3000000000>19/100\\
 3&349/1500&1097307737/1171875000000>9/10000.
\end{array}
\]
Therefore $E_j>0$ on $[0,b_j]$, which contains the required band.
Combining this with \eqref{eq:universal-Xi-bound} gives
\[
 \theta_j<\frac{(1+s\lambda_j(s))^2}
 {(1-s^2\lambda'_j(s))(1+23s/12-s^2/4)}<1.
\]
This proves \cref{lem:theta}.

\section{Proof of Lemma  \ref{lem:curvature}}
\label{app:curvature}

\subsection{Convention for the displayed polynomial identities}
\label{app:positive-products}
For a rational function in one variable, $\operatorname{pn}$ denotes
its primitive integer numerator: reduce to lowest terms, clear
rational coefficient denominators, normalize the denominator to have
positive leading coefficient, and divide the numerator by its positive
integer content. In every application below the reduced denominator
is positive on the stated domain. Thus taking $\operatorname{pn}$
preserves the sign of the function. All coefficient lists are in
\emph{ascending} order.

For example, an identity
\[
 P(h)=\sum_{k=0}^n A_k h^k(1-h)^{n-k},\qquad A_k>0,
\]
proves $P>0$ on $[0,1]$ directly. Likewise a positive expansion
$\sum B_k(y-1)^k(2-y)^{n-k}$ proves positivity on $[1,2]$.
We give the actual integer coefficients of the identities, rather
than the output or subdivision count of a verifier. All identities
can be checked by multiplication and coefficient comparison.

\subsection{Monotonicity of \texorpdfstring{$R$ in \cref{lem:curvature}}{R in the curvature lemma}}
\label{app:curvature-cert}
For the $j$th quadratic define
\[
 M_j(h)=\frac{h}{1+h\lambda_j(h)},\quad
 S_j(h)=(1+2h^2)M_j'(h)-(1-h^2)hM_j''(h),\quad
 T_j(h)=S_j(h)-\frac{1-h^2}{h}S_j'(h).
\]
The apparent singularity at $h=0$ in $T_j$ is removable.
Writing $x=2t-1$ and $h=\sqrt{1-x^2}$ gives
$R_j(t)=xS_j(h)$ and $R_j'(t)=2T_j(h)$.
The denominator of $T_j$ is a positive constant times
$(1+h\lambda_j(h))^4$. Set $N_j(h)=\operatorname{pn}_hT_j(h)$.
Then
\begin{equation}\label{eq:R-positive-identity}
 N_j(h)=\sum_{k=0}^{10}a_{jk}h^k(1-h)^{10-k},\CASmark{30}
\end{equation}
with the following strictly positive integer coefficient vectors.
{\footnotesize\[
\begin{aligned}
\boldsymbol a_1={}&(102950,\ 790200,\ 2422525,\ 3707000,\ 3510429,\ 5380891,\ 11700069,\ 14790276,\ 9253764,\\
&\quad 2496420,\ 170100).\end{aligned}
\]}
{\footnotesize\[
\begin{aligned}
\boldsymbol a_2={}&(170749000,\ 1326362000,\ 4118284500,\ 6439624000,\ 6350391647,\ 9528566146,\ 19763726778,\\
&\quad 24659631592,\ 15552488988,\ 4346743214,\ 339067289).\end{aligned}
\]}
{\footnotesize\[
\begin{aligned}
\boldsymbol a_3={}&(217585600,\ 1773344000,\ 5933787200,\ 10697785600,\ 13192478082,\ 18236780616,\\
&\quad 29118442751,\ 32571534052,\ 20159814580,\ 5901969500,\ 554963955).\end{aligned}
\]}

Thus $T_j>0$ on $[0,1]$, proving $R_j'(t)>0$.
The formula for $S_j$ gives $S_j(0)=1$, so the endpoint values
are $R_j(0+)=-1$ and $R_j(1-)=1$.

\subsection{The left-hand sign for unimodality}
\label{app:unimodality-identities}
\noindent\textbf{Result proved:} \eqref{eq:F1-left}. 
Let $y=\sqrt{t/(1-t)}$ and
\[
 R_j(y)=\frac{y^2-1}{1+y^2}
       S_j\!\left(\frac{2y}{1+y^2}\right),\qquad
 f_{1,j}=(1+R_j)\frac{3(4-y^2)}{4y(1+y^2)}-R_j'.
\]
All rational denominators have known positive sign. Indeed the
polynomial
\[
 D_j(y)=(1+y^2)^3+2L_jy(1+y^2)^2
              +4U_jy^2(1+y^2)+8V_jy^3
\]
has the positive coefficient list \CASmark{31}
$(1,2L_j,3+4U_j,4L_j+8V_j,3+4U_j,2L_j,1)$.
The denominators in question are products of positive constants,
powers of $D_j$, $y$, and $1+y^2$.

The primitive numerator of $f_{1,j}$ has one factor $y$.
After removing it, write $L_j^*(y)=\operatorname{pn}_y(f_{1,j})/y$.
Its degree is $24$, and
\begin{equation}\label{eq:Fleft-positive-identity}
 L_j^*(y)=\sum_{k=0}^{24}b_{jk}y^k(1-y)^{24-k}.\CASmark{32}
\end{equation}
The positive coefficient vectors are
{\footnotesize\[
\begin{aligned}
\boldsymbol b_1={}&(5147500,\ 162249200,\ 2243294899,\ 19081994486,\ 114165246654,\ 513625348436,\\
&\quad 1801790426684,\ 5025684923008,\ 11247494070087,\ 20230772257488,\ 29153975303672,\\
&\quad 33632124470288,\ 32037437915328,\ 29350480947232,\ 34296655804448,\ 49926284114112,\\
&\quad 67722924709248,\ 73982991859200,\ 62747068884480,\ 40776126796800,\ 19930758912000,\\
&\quad 7074107136000,\ 1708335360000,\ 246965760000,\ 15552000000).\end{aligned}
\]}
{\footnotesize\[
\begin{aligned}
\boldsymbol b_2={}&(42687250000,\ 1327061228000,\ 18296835660504,\ 155643131505112,\ 931535960474763,\\
&\quad 4190386680733116,\ 14686229654642150,\ 40879395632984356,\ 91101686559155447,\\
&\quad 162350158757049304,\ 228840484612582232,\ 249420675715882976,\ 203882819267955512,\\
&\quad 129686961465271552,\ 106395749011442176,\ 181002086406024832,\ 309157930752771472,\\
&\quad 389072613763529472,\ 358263321906074240,\ 244502363714628864,\ 122985578865543040,\\
&\quad 44267966824694272,\ 10694365217947136,\ 1519231014918144,\ 90761024451328).\end{aligned}
\]}
{\footnotesize\[
\begin{aligned}
\boldsymbol b_3={}&(33997750000,\ 1044954844000,\ 14471910282824,\ 124304340344168,\ 753019166564705,\\
&\quad 3436919155727000,\ 12270053578527874,\ 35019743015840888,\ 80869885291382180,\\
&\quad 151865479135353960,\ 231825788668311704,\ 286515881543217508,\ 286225832538423158,\\
&\quad 236883677748058876,\ 183341429594299064,\ 169428346828570568,\ 193234528036138243,\\
&\quad 211015300107527320,\ 186395320683924032,\ 125647295959293168,\ 62687177842348728,\\
&\quad 22220407704404000,\ 5192452094847360,\ 685201689268544,\ 33600967150128).\end{aligned}
\]}

This proves $f_{1,j}>0$ for $0<y\le1$.
\subsection{A coarse curvature bound in place of the central derivative test}
\label{app:central-curvature}
\noindent\textbf{Result proved:} \eqref{eq:Tprime-bound}. Together
with \eqref{eq:F1-left} and $R'>0$, it proves the unimodality
assertion in \cref{lem:curvature} by the positive cubic in
\eqref{eq:F-decreasing}. In particular, no sign assertion about
$f'_{1,j}$ is required.

Let $D_j(h)=1+h\lambda_j(h)$ and use the already defined
$S_j,T_j$. By eqref{eq:R-positive-identity}, the expression
\[
 N_j^0(h):=D_j(h)^4 T_j(h)
\]
is a polynomial of degree at most ten. It can be formed without any
rational-function differentiation as follows:
\begin{align*}
 A_j^0&=D_j-hD_j',\\
 B_j^0&=(1+2h^2)A_j^0D_j
       -(1-h^2)h\bigl((A_j^0)'D_j-2A_j^0D_j'\bigr),\\
 N_j^0&=B_j^0D_j-
   \frac{1-h^2}{h}\bigl((B_j^0)'D_j-3B_j^0D_j'\bigr).
\end{align*}
The numerator in the last quotient is divisible by $h$.
These formulas follow from $M'=A_j^0/D_j^2$,
$S_j=B_j^0/D_j^3$ and the definition of $T_j$.
Now define the degree-fifteen polynomial
\begin{equation}\label{eq:coarse-curvature-polynomial}
 P_j(h)=3D_j(h)^5-D_j(h)(N_j^0)'(h)
                         +4D_j'(h)N_j^0(h).\CASmark{33}
\end{equation}
It satisfies the identity
\[
 3-T_j'(h)=P_j(h)/D_j(h)^5.
\]
It is enough to prove $P_j>0$ on $[4/5,1]$.

The needed coefficient arithmetic can be shared by all three
kernels. 
Since $P_j$ has degree at most $15$, its Taylor
expansion at $h=1$ is exact. Setting $h=1-u$
gives
\[
\frac{P_j(1-u)}{P_j(1)}
=\sum_{k=0}^{15}
c_{jk} u^k,
\]
where
\[\label{eq:normalized-curvature-coefficients}
c_{jk}=\frac{(-1)^kP_j^{(k)}(1)}
{k!P_j(1)}
=\frac{(-1)^k}{P_j(1)}
\sum_{\ell=k}^{15}\binom{\ell}{k}
[h^\ell]P_j(h). \CASmark{34}
\]
Substituting the three quadratics of \cref{tab:source} into
\eqref{eq:coarse-curvature-polynomial} gives $P_j(1)>70$ and the
following common rational lower bounds $c_{jk}\ge l_k$:
\begin{center}\small
\begin{tabular}{@{}c|rrrrrrrr@{}}\toprule
$k$&0&1&2&3&4&5&6&7\\\midrule
$l_k$&1&$-7$&14&$-16$&$-13$&0&$-13$&0\\\midrule
$k$&8&9&10&11&12&13&14&15\\\midrule
$l_k$&0&$-11/10$&0&$-3/100$&0&$-1/500$&0&$-1/20000$\\\bottomrule
\end{tabular}
\end{center}
For $0\le u\le1/5$, absorb the negative terms of degree at least
three into the coefficient of $u^2$. The resulting coefficient is
\[
 14-\frac{16}{5}-\frac{13}{5^2}-\frac{13}{5^4}
 -\frac{11}{10\cdot5^7}-\frac{3}{100\cdot5^9}
 -\frac1{500\cdot5^{11}}-\frac1{20000\cdot5^{13}}
 >\frac{41}{4}.
\]
Consequently,
\begin{equation}\label{eq:coarse-curvature-quadratic}
 \frac{P_j(1-u)}{P_j(1)}
 \ge1-7u+\frac{41}{4}u^2\ge\frac1{100}>0
 \qquad(0\le u\le1/5).\CASmark{35}
\end{equation}
The last quadratic is decreasing on this interval and has value
$1/100$ at $u=1/5$. This proves \eqref{eq:Tprime-bound} for all three
kernels simultaneously. The middle-interval argument in the proof
of \cref{lem:curvature} now establishes the unique sign change of
$W_\vartheta'$ for every $0\le\vartheta<1$.

\section{Mathematica verification index}\label{app:CAS-index}
The numbered clover symbols identify finite calculations independently
reconstructed in the companion file
\texttt{NoiseStability\_CAS\_Verification.wl}. The code prints
\texttt{PASS/FAIL} for every indexed check and includes the matching
\texttt{CAS-nn} identifier alongside the original equation or table label.
One symbol may cover the corresponding three kernel rows or several
arithmetic steps within the same displayed claim. 
These computations verify the stated algebraic identities,
coefficient inequalities, and fixed-value bounds.
The code does not independently verify
the external hypothesis in \cref{ass:input}.

{\small\setlength{\tabcolsep}{4pt}
\begin{longtable}{@{}p{.11\textwidth}p{.26\textwidth}p{.54\textwidth}@{}}
\toprule Mark & Formula / data & Exact scope of the companion calculation\\\midrule
\endfirsthead
\toprule Mark & Formula / data & Exact scope of the companion calculation\\\midrule
\endhead
\bottomrule\endfoot
\CASmark{01} & \eqref{eq:thresholds} & Exact overlap of the two correlation thresholds \\[2pt]
\CASmark{02} & \eqref{eq:Phi-curvature} & Normalized scalar kernel and its second derivative \\[2pt]
\CASmark{03} & \eqref{eq:small-scalar} & Weighted AM--GM coefficients in the small-noise majorant \\[2pt]
\CASmark{04} & \eqref{eq:small-constant} & Exact rational bound for the small-noise constant \\[2pt]
\CASmark{05} & Table~\ref{tab:low} & Three rows of scalar parameters and mixed-moment side conditions \\[2pt]
\CASmark{06} & \eqref{eq:contact-identity} & Symmetric contact algebra for K \\[2pt]
\CASmark{07} & \eqref{eq:Mprime} & Differentiation of the Hellinger auxiliary kernel M \\[2pt]
\CASmark{08} & \eqref{eq:theta} & Algebraic simplification of the contact coefficient ratio \\[2pt]
\CASmark{09} & \eqref{eq:F-decreasing} & Formula for Q double prime and the positive cubic \\[2pt]
\CASmark{10} & \eqref{eq:E-factor} & Factorization of the contact residual second derivative \\[2pt]
\CASmark{11} & \eqref{eq:local-complex-disk} & Rational constants in the complex-disc estimate \\[2pt]
\CASmark{12} & \eqref{eq:local-coefficient-bounds}, Appendix~\ref{app:taylor-data} & All 14 exact Taylor coefficients and rational coefficient bounds \\[2pt]
\CASmark{13} & Appendix~\ref{app:local-interval} & Exact Cauchy-tail rational constant \\[2pt]
\CASmark{14} & \eqref{eq:local-single-endpoint} & Certified rational enclosure at the local endpoint \\[2pt]
\CASmark{15} & \eqref{eq:W1-factor} & Exact factorization of the degree-five polynomial majorant \\[2pt]
\CASmark{16} & Appendix~\ref{app:W1-proof} & Five Rademacher even-moment identities (orders 2--10) \\[2pt]
\CASmark{17} & \eqref{eq:W1-one-variable-reduction} & Moment-reduction coefficients and exact slack identity \\[2pt]
\CASmark{18} & \eqref{eq:W1-tangent} & Exact cubic tangent expansion and positive constants \\[2pt]
\CASmark{19} & \eqref{eq:majorant-derivatives} & Three scalar residual derivative identities \\[2pt]
\CASmark{20} & \eqref{eq:majorant-shape-points} & Fixed algebraic eighth-root shape comparisons \\[2pt]
\CASmark{21} & \eqref{eq:scalar-point-values}, Table~\ref{tab:scalar-points} & Nine rigorous fixed-point residual and slope bounds \\[2pt]
\CASmark{22} & \eqref{eq:Bpp-analytic} & Exponential exponent derivatives and rational series bound \\[2pt]
\CASmark{23} & Appendix~\ref{app:expectation-interval} & Psi double-prime and Delta concavity comparisons \\[2pt]
\CASmark{24} & Table~\ref{tab:endpoint-gaps} & Six rigorous lower bounds for expectation-gap endpoints \\[2pt]
\CASmark{25} & \eqref{eq:theta-bounds} & Three kernel derivative and positivity comparisons \\[2pt]
\CASmark{26} & Appendix~\ref{app:theta-cert} & Symmetric change of variables s=2/H and Xi=HJ/K \\[2pt]
\CASmark{27} & \eqref{eq:universal-ratio-factorization} & Exact positive degree-ten universal factorization \\[2pt]
\CASmark{28} & \eqref{eq:universal-Xi-bound} & Exact rational-difference identity for the Xi upper bound \\[2pt]
\CASmark{29} & \eqref{eq:quintic-data} & Three quintic coefficient lists, derivative signs and endpoints \\[2pt]
\CASmark{30} & \eqref{eq:R-positive-identity} & Three degree-ten primitive-numerator product identities \\[2pt]
\CASmark{31} & Appendix~\ref{app:unimodality-identities} & Positive coefficient lists for three $D_j$ denominator polynomials \\[2pt]
\CASmark{32} & \eqref{eq:Fleft-positive-identity} & Three degree-24 primitive-numerator product identities \\[2pt]
\CASmark{33} & \eqref{eq:coarse-curvature-polynomial} & Three coarse-curvature polynomial identities \\[2pt]
\CASmark{34} & \eqref{eq:normalized-curvature-coefficients} & Forty-eight normalized rational coefficient bounds \\[2pt]
\CASmark{35} & \eqref{eq:coarse-curvature-quadratic} & Exact coefficient excess and endpoint quadratic check \\[2pt]
\end{longtable}}

\end{document}